\documentclass[twocolumn,pra,amsmath,amssymb]{revtex4}

\begin{document}

\title{ Cluster States From Heisenberg Interaction}

\author{Massoud Borhani}

\author{Daniel Loss}

\affiliation{Department of Physics and Astronomy, University of Basel, Klingelbergstrasse
82, CH-4056 Basel, Switzerland}

\begin{abstract}
We show that a special type of entangled states, cluster
states, can be created with  Heisenberg interactions and local
rotations in 2$d$ steps where $d$
is the dimension of the lattice. We find that, by
tuning the coupling strengths, anisotropic exchange interactions can
also be employed to create cluster states. Finally, we propose 
electron spins in quantum dots as a possible  realization of a one-way
quantum computer based on cluster states.
\end{abstract}

\maketitle

\section{Introduction}
Entanglement plays a crucial role in quantum information
processing \cite{niel}. Quantum algorithms (in particular, Shor's
algorithm, to find the prime factors of an n-bit integer) exploit entanglement
to speed up computation. In addition, quantum communication protocols
use  entangled states as a medium to send information
through quantum channels.
However, creating entangled states is a great challenge for
both theoretical and experimental physicists. Recently, Briegel
 and Raussendorf \cite{br} introduced a special
kind of entangled states, the so-called cluster states, 
 which can be created via an Ising Hamiltonian \cite{hu}.
These states are eigenstates of certain correlation operators
(see Eqs. (\ref{Cdef}) and (\ref{Cdef1}) below). It
has been shown that via cluster states, one can implement a quantum
computer on a lattice of qubits.
In this proposal, which is known as ``one way quantum computer'', information
is  written onto the cluster, processed, and read out from the cluster
by one-qubit measurements \cite{rauss}. In other words, all types of
quantum circuits and quantum gates can be implemented on the lattice of
qubits by single-qubit measurements only. The entangled state of the cluster
thereby serves as a universal resource for any quantum computation.
However, in this model, cluster states are created with an Ising interaction,
which maybe difficult to realize, in particular in a solid state system.
 Here, we propose an alternative
way to create the same states with a  Heisenberg interaction
(isotropic exchange interaction), but in several
steps, where the number of steps depends on the dimension of the
 lattice of cubic symmetry.
Furthermore, we consider some deviations from the Heisenberg Hamiltonian
due, for example, to  lattice asymmetry, 
and  obtain the same cluster state by tuning the exchange coupling
strengths. It turns out that if these coupling strengths  satisfy
certain conditions, which can be tuned  experimentally, 
we can obtain a cluster state, up to an
overall phase. Following Ref. \cite{loss}, we propose a lattice of 
electron spins in  quantum dots as a possible realization of
this scheme in solid-state systems. In this system, electron spins in
 nearest-neighbor quantum dots are coupled
via a Heisenberg exchange interaction.

This paper is organized as follows: Section II is devoted to a brief
introduction to cluster states. In Section III we introduce an alternative
way to create cluster states. Section III.C considers the anisotropic
Heisenberg interaction between qubits on a lattice and how to get
cluster states via this interaction. Finally, in Section IV, we propose
electron spins in quantum dots as a physical realization of this proposal.\\

\section{Cluster States}

A cluster state \cite{br} is an entangled state
which has special features suitable for implementing a quantum computer
on an array of qubits. According to this scheme, we can obtain a cluster state
by applying an Ising Hamiltonian ($\hbar = 1$)
 \begin{eqnarray}
H= g(t)
\sum_{<a,\, a'>}\frac{1-\sigma_{z}^{(a)}}{2}
\frac{1-\sigma_{z}^{(a')}}{2},\label{HIsing}
\end{eqnarray}
 on a special kind of initial state. 
Here, $\sigma_{i}^{(a)}$, $i \in \{x, y, z\}$ are 
Pauli matrices at lattice site $a$ and $<a,\, a'>$
denotes that $a'$ is the nearest neighbor of $a$ . Furthermore, 
$g(t)$ allows for a possible overall time dependence.
To be specific, consider a qubit chain (see Figure 1.a) prepared initially
in a product state 
$|\phi_{0}\rangle=\bigotimes_{a}|+\rangle_{a}$,
where index $a$ refers to the sites of the qubits
and $|+\rangle_{a}$ is  eigenstate of $\sigma_{x}^{(a)}$ 
with eigenvalue $1$.
The time evolution operator for the qubit chain is then given by
\begin{eqnarray}
U(\theta)=exp\,\,(-i\,\theta\sum_{a}\frac{1-\sigma_{z}^{(a)}}{2}
\frac{1-\sigma_{z}^{(a+1)}}{2})\,,
\end{eqnarray}
with $\theta=\int g(t)dt$. From now on we assume that $\theta=\pi$ \cite{br}.
Because the terms in the Ising Hamiltonian (\ref{HIsing}) mutually commute, we
can decompose the evolution operator $U(\pi)$ into two-particle operators
as follows,
\begin{eqnarray}
U\,\equiv\,\, U(\pi) & = & \prod_{a}U^{(a, a+1)} \;,\label{U}\\
U^{(a, a+1)} & = & \frac{1}{2}\,(\,1+\sigma_{z}^{(a)}+ 
\sigma_{z}^{(a+1)}- \sigma_{z}^{(a)}
\sigma_{z}^{(a+1)}\,).\label{U1}\;\;\;\;\;\;\;\;
\end{eqnarray}
Therefore, $U$ is a product of two-qubit conditional phase gates \cite{niel}.
More generally we can define the cluster states as the eigenstates
of the following correlation operators 
 \begin{eqnarray}
K^{(a)}\,\left|\,\phi_{\{\kappa\}}\,\right>_{\mathcal{C}}\, & = & 
\,(-1)^{\kappa_{a}}\,\left|\,\phi_{\{\kappa\}}\,\right>_{\mathcal{C}} \;,\label{Cdef}\\
K^{(a)}\, & \equiv & \,\sigma_{x}^{(a)}\bigotimes_{b\in\,\,
  nbgh(a)}\sigma_{z}^{(b)}\label{Cdef1}\,\,,
\label{Coperat}
\end{eqnarray}
 with $\kappa\,\in\,\{0,1\}\,$. A cluster state is completely
specified by the eigenvalue equation (\ref{Cdef}) and it can be shown
\cite{rbb} that all states $\left|\,\phi_{\{\kappa\}}\,\right>_{\mathcal{C}}$
are equally suitable for computation. For simplicity we put $\kappa=0$
for all lattice sites. The detailed proof of the above assertions
and properties of cluster states, especially their application in
implementing a one way quantum computer, have been given in
Refs. \cite{rauss} and \cite{rbb}.
We note that in one dimension a cluster is a qubit chain with nearest
neighbor interaction.
However, in more than one dimension, the cluster does not have a regular shape.
In the latter case, qubits can be arranged
in a multi-dimensional square  lattice such that only some of the
lattice sites are occupied by qubits. A  cluster is then defined as
a set of  qubits where any two qubits are connected by a sequence of
neighboring sites that are occupied by a qubit.

\section{Cluster States From Heisenberg Interaction}

Cluster states are produced through Ising interactions. 
However, an ideal Ising interaction
is  difficult to obtain in nature especially in a solid state
environment. So, how can such states be created? The usual spin-spin
interaction is (nearly) isotropic in spin space and is described by
the Heisenberg Hamiltonian \cite{ash}, 
\begin{eqnarray}
H  &=&  -J\,\sum_{<ij>}
S_{x}^{(i)}S_{x}^{(j)}+S_{y}^{(i)}S_{y}^{(j)}+S_{z}^{(i)}S_{z}^{(j)}\label{H}\,\,,\\
\vec{S}^{(i)} &=& (S_{x}^{(i)}, S_{y}^{(i)}, S_{z}^{(i)}) = {1 \over 2}
\vec{\sigma}^{(i)}\;\;\;\;\;\;\;(\hbar =1),
\end{eqnarray}
 where $\vec{S}^{(i)}$ and  $\vec{S}^{(j)}$ are spin-${1 \over 2}$ operators
 at lattice sites $i$ and $j$, and $J$ is the
exchange coupling constant,
which is assumed to be constant for all spin pairs and is
positive (negative)
 for ferromagnetic (antiferromagnetic) coupling. Next
we describe a method to create cluster states via Heisenberg instead
of Ising interaction. We start with one dimension and
then generalize to higher dimensions.\\

\subsection{One-Dimensional Case}

Recall that all operators $U^{(a, a+1)}$ in $U$ (Eqs. (\ref{U}) and
(\ref{U1}) above) mutually commute and they can therefore be applied
in arbitrary order, i.e. at the same or different times. 
(see Fig. 1.a). Suppose we have
a one-dimensional N-qubit chain where all qubits are prepared in the 
$\left|+\right>$ state. The initial state of the cluster is then
(as before)\,\, $\bigotimes_{a\in\,\mathcal{C}}\left|+\right>_{a}$\,\,,
and the index $a$ refers to the lattice site. The idea is to
apply first the sequence $U^{(1, 2)}\, U^{(3, 4)}\, U^{(5, 6)}\dots$\,
, and then in a second step, 
 the sequence $U^{(2, 3)}\, U^{(4, 5)}\, U^{(6, 7)}\dots$\,\,.
In other words, first we let qubits 1-2, 3-4, 5-6, ... interact with each
other, and then qubits 2-3, 4-5, 6-7, ... (Fig. 1.b). We obtain the same
result (\ref{U}), but now we have entangled the qubits in our chain
\textit{pairwise} in each step.
This means that each qubit is entangled with only one of its nearest
neighbors in each step. In one dimension, there are two nearest neighbors
for each qubit, thus we entangle our chain in two steps.

We note that $U^{(a, a+1)}$\,, given by Eq.(\ref{U1}),
describes a conditional phase shift. On the other hand, 
in Ref. \cite{loss} it was shown that this evolution operator can also be realized
with a Heisenberg Hamiltonian (obtained e.g.  via a Hubbard model) and local one-qubit
rotations (see also next section).
Therefore, the problem of generating a cluster state with a Heisenberg
interaction has been solved provided in each step each qubit interacts
with only one of its nearest neighbors.

\subsection{Higher Dimensions}

In two dimensions, the minimum number of steps increases to four 
in a two dimensional square lattice.
 In general for a $d$-dimensional cubic lattice, the minimum
number of steps required is $2d$. (Note that cluster states are only defined
on lattices with cubic symmetry. See also  the last paragraph  in Section II).

However, in dimensions higher than one, there is no regular shape
for an arbitrary cluster. How then, can we obtain cluster states with
just $2d$ steps? There may be several optimal ways to do this but
we mention only one. For simplicity, consider a two dimensional cluster
and suppose that this cluster can be contained within a rectangle
of $n$ rows and $m$ columns. Now, entangle all qubits in the cluster
within each of these $n$ rows independently (recall that each row
requires two steps to be entangled). Then, do the same for the $m$
columns. There is no need to worry about the qubits which are within the
rectangle but not part of the cluster, since they are excluded automatically
if we do not entangle them to their nearest neighbors. The idea is
the same for $d\,=\,3$  cubic lattice, except that we would need 6 steps to entangle
the cluster. 

\setlength{\unitlength}{0.030in}
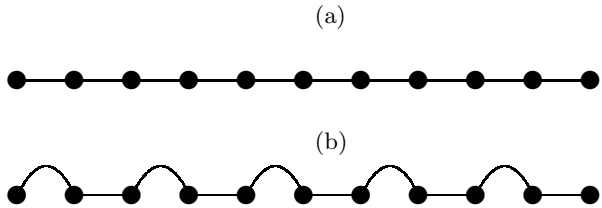
\begin{figure}
 \begin{picture}(100,40)
 \put(100,30){\circle*{3}}
 \put(90,30){\circle*{3}}
 \put(80,30){\circle*{3}}
 \put(70,30){\circle*{3}}
 \put(60,30){\circle*{3}}
 \put(50,30){\circle*{3}}
 \put(40,30){\circle*{3}}
 \put(30,30){\circle*{3}}
 \put(20,30){\circle*{3}}
 \put(10,30){\circle*{3}}
 \put(0,30){\circle*{3}}
 \put(52,40){(a)}
 \put(90,30){\line(1,0){10}}
 \put(80,30){\line(1,0){10}}
 \put(70,30){\line(1,0){10}}
 \put(60,30){\line(1,0){10}}
 \put(50,30){\line(1,0){10}}
 \put(40,30){\line(1,0){10}}
 \put(30,30){\line(1,0){10}}
 \put(20,30){\line(1,0){10}}
 \put(10,30){\line(1,0){10}}
 \put(0,30){\line(1,0){10}}

\put(100,10){\circle*{3}}
 \put(90,10){\circle*{3}}
 \put(80,10){\circle*{3}}
 \put(70,10){\circle*{3}}
 \put(60,10){\circle*{3}}
 \put(50,10){\circle*{3}}
 \put(40,10){\circle*{3}}
 \put(30,10){\circle*{3}}
 \put(20,10){\circle*{3}}
 \put(10,10){\circle*{3}}
 \put(0,10){\circle*{3}}
 \put(52,18){(b)}
 \put(90,10){\line(1,0){10}}
 \qbezier(80,10)(85,20)(90,10)
 \put(70,10){\line(1,0){10}}
 \qbezier(60,10)(65,20)(70,10)
 \put(50,10){\line(1,0){10}}
 \qbezier(40,10)(45,20)(50,10)
 \put(30,10){\line(1,0){10}}
 \qbezier(20,10)(25,20)(30,10)
 \put(10,10){\line(1,0){10}}
 \qbezier(0,10)(5,20)(10,10)
  \end{picture}
 \caption{(a) A one-dimensional cluster (a qubit chain). The
   connecting lines represent
the interaction between nearest neighbors.
(b) An alternative way to entangle a one-dimensional cluster. The qubits which are
connected by straight lines are entangled in the first step and
those connected by semicircles are entangled in the subsequent step. }
\end{figure}
                                                                
\subsection{Anisotropic Heisenberg Hamiltonian}

We do not consider the most general form of an anisotropic Heisenberg
model since it is beyond the scope of this work. Here we introduce
a special case, known as $symmetric$ anisotropic Heisenberg
model (SAH) which does not include the cross-spin terms. It has the
following form in one dimension
 \begin{eqnarray}
 H_{SAH}  &=& \sum_a H_{SAH}^{(a, a+1)}\;,\\
H_{SAH}^{(a, a+1)} &=& \alpha(t)\, S_{x}^{(a)}S_{x}^{(a+1)} \nonumber\\
& & +\beta(t)\,
S_{y}^{(a)}S_{y}^{(a+1)}+\gamma(t)\, S_{z}^{(a)}S_{z}^{(a+1)}.\label{SAH}\;\;\;\;\;\;\;
\end{eqnarray}
 This situation occurs for example, when our lattice does not have
enough symmetry to use the isotropic interaction. However, \begin{eqnarray}
\left[\, S_{p}^{(a)}S_{p}^{(a+1)}\,,\, S_{q}^{(a)}S_{q}^{(a+1)}\,\right] & = & 0\,\,,\\
\forall \; p, q   =  x, y, z\,\,.\nonumber\;\;\;\;\;\;\;
\end{eqnarray}
 Therefore, these three terms in the Hamiltonian mutually commute
and consequently when we write the unitary evolution operator for two adjacent
qubits, $U_{SAH}^{\,\,\,(a, a+1)}$, it can be decomposed into
three unitary operators. The order of application of these three operators
does not matter 
\begin{eqnarray}
U_{SAH}^{\,\,\,(a, a+1)} & = & U_{xx}^{(a, a+1)}\, U_{yy}^{(a, a+1)}\,
U_{zz}^{(a, a+1)}\;,\\
U_{xx}^{(a, a+1)} & = & exp\,\,(\,-i\, J_{xx}\, S_{x}^{(a)}\, S_{x}^{(a+1)})\;,\\
U_{yy}^{(a, a+1)} & = & exp\,\,(\,-i\, J_{yy}\, S_{y}^{(a)}\, S_{y}^{(a+1)})\;,\\
U_{zz}^{(a, a+1)} & = & exp\,\,(\,-i\, J_{zz}\, S_{z}^{(a)}\,
S_{z}^{(a+1)})\,\,.
\end{eqnarray}
 Now, according to our alternative method to create cluster states,
if the coefficients $\alpha$, $\beta$ and $\gamma$ satisfy the
following conditions, 
 \begin{eqnarray}
J_{xx}=\int\alpha(t)\, dt & = & 4n\pi \;,\label{cond1}
\;\;\;\;\;\;\;\;\;\;\;\;\;\;\;\;\;\;\;\;\\
J_{yy}=\int\beta(t)\, dt & = & 4m\pi \;,\label{cond2}\;\;\;\;\;\;\\
J_{zz}=\int\gamma(t)\, dt & = &  (2k+1)\pi \;,\label{cond3}
\end{eqnarray}
 where $n$, $m$, and $k$ are arbitrary integers,  Then
 $U_{xx}^{(a, a+1)}$
and $U_{yy}^{(a, a+1)}$ are just \textit{unity} operators (up to
a minus sign) and do not affect the initial state \cite{ising}.
 If we could tune these coefficients properly in our lattice,
we would get the same cluster states, up to some local 
 (single-qubit) operations. The crucial point is that 
$U_{SAH}^{\,\,\,(a, a+1)}$ and $U_{SAH}^{\,\,\,(a+1, a+2)}$
do not commute and thereby, we can $not$ decompose $U_{SAH}$, the total
evolution operator of the cluster with an SAH interaction,
 into two-qubit
evolution operators. This is why we need at least two steps to entangle
the chain.\\
 In general, when the Hamiltonian includes cross-spin terms, (the
asymmetric anisotropic Heisenberg model (AAH)), the problem can not
be solved exactly because the terms in the AAH Hamiltonian do not
mutually commute. There is still a hope of solving this problem if
we have the following interaction between spins \cite{cycle}: 
\begin{eqnarray}
H_{AAH}^{(a, a+1)} &=& \alpha'(t)\, S_{x}^{(a)}S_{y}^{(a+1)}\nonumber\\
& & +\beta'(t)\,
S_{y}^{(a)}S_{x}^{(a+1)}+\gamma'(t)\, S_{z}^{(a)}S_{z}^{(a+1)}.\;\;\;\;\;
\end{eqnarray}
Again, the terms in this Hamiltonian mutually commute and we can decompose
the two-qubit evolution operator like above.
However, this Hamiltonian is related to the previous Hamiltonian~(\ref{SAH}) via a
single-qubit unitary transformation (through $\pi/2$-rotation of  one of the 
spins about the $z$-axis) and therefore, both have  the same structure.
In the end, we emphasize that the basic cornerstone of this
method is that in each step, each qubit can interact with $only$
one of its nearest neighbors. Generalizing the above method to higher
dimensions is straightforward (see previous section).
 Therefore we have shown that the problem
of creating cluster states with more realistic interaction models
other than Ising, can be solved exactly.\\

\section{Physical Realization Of The Model}

In Refs. \cite{loss}, \cite{bur} and \cite{awschalom},  a detailed scenario
has been proposed  
for how quantum computation may be achieved
in a coupled quantum dots system. In this proposal, a qubit is realized
as the spin of the excess electron on a single-electron quantum dot. 
A  mechanism has been proposed there for two-qubit quantum-gate
operation that operates by a purely electrical gating of the tunneling
barrier between neighboring quantum dots, rather than by spectroscopic
manipulation as in other models. Consider two quantum dots which are
labeled by ``$1$'' and ``$2$'' and 
coupled to each other via exchange interaction (see below).
If the barrier potential is ``high'',
tunneling is forbidden between dots, and the qubit states are held
stably without evolution in time ($t$). If the barrier is pulsed
to a ``low'' voltage, the usual physics of the Hubbard model \cite{ash}
says that the spins will be subject to a transient Heisenberg
coupling,                
\begin{equation}
H = J(t) \vec{S}^{(1)} \cdot \vec{S}^{(2)} \;\;,\label{Exc}
\end{equation}
 where $J(t)$ is the time-dependent exchange
constant which is produced by the turning on and off of the tunneling
matrix element \cite{loss,bur}.    

 For instance, a quantum XOR gate is obtained by a simple sequence
of operations \cite{loss}:
 \begin{equation}
U_{XOR}=e^{i\frac{\pi}{2}S_{z}^{(1)}}e^{-i\frac{\pi}{2}S_{z}^{(2)}}U_{sw}^{\frac{1}{2}}e^{i\pi
 S_{z}^{(1)}}U_{sw}^{\frac{1}{2}}\,\,,\label{XOR}
\end{equation}
 where $U_{sw}$ is a swap gate, created in this model via Heisenberg
interaction, and $e^{i\pi S_{z}^{(1)}}$ etc. are single-qubit operations
only, which can be realized  by applying local Zeeman interaction.
(It has been established that XOR along with single-qubit operations
may be assembled to do any quantum computation \cite{bar}.) Note that
the XOR of Eq.\ (\ref{XOR}) is given in the basis where it has the
form of a conditional phase-shift operation; the standard XOR is obtained
by a simple basis change for qubit {}``\,2\,''\,.
According to Eq. (\ref{XOR}), we need $5$ steps to realize an XOR
gate. However,
in Ref. \cite{Smolin}  it has been shown that for a certain choice 
of system parameters (for example, opposite direction of the local $B$ fields), 
we can generate an XOR gate in one step.
The crucial observation now is that the XOR operation can be written 
as  \cite{loss} $U_{XOR}={\frac{1}{2}}+S_{z}^{(1)}+S_{z}^{(2)}-2S_{z}^{(1)}S_{z}^{(2)}$,
which has exactly the same form as $U^{(a, a+1)}$ in Eq. (\ref{U1}).
In other words, we can generate the operation $U^{(a, a+1)}$ 
(and thus the cluster states) with the Heisenberg interaction
 as described e.g. by the sequence in Eq. (\ref{XOR}).
We finally note that an alternative way to achieve the XOR operation is given by  \cite{loss}
 $U_{XOR}=e^{i{\pi}S_{z}^{(1)}}\, U_{sw}^{-\frac{1}{2}}\,
 e^{-i{\frac{\pi}{2}}S_{z}^{(1)}}\, U_{sw}\,
 e^{i{\frac{\pi}{2}}S_{z}^{(1)}}\, 
U_{sw}^{\frac{1}{2}}$.
This form has the potential advantage that the single qubit operations
involve only spin 1.

The mechanisms described above
for performing gate operations with spin qubits are independent of the
details of the pulse shape $P(t)$, where $P$ stands for the exchange
coupling $J$ or the Zeeman interaction.  It is only the value of the
integral
$\int_0^{\tau} P(t) dt$ (mod $2\pi$) which determines the quantum gate
action. This is true provided that the parameters $P(t)$ are switched
adiabatically, guaranteeing the validity of the effective Hamiltonian
Eq.~(\ref{Exc}).
The unwanted admixture of a state with double occupation of a dot in the
final state
is found to be tiny if a suitable pulse is used and the adiabaticity
criterion is fulfilled \cite{Schl,Req}.

We  note that as long as an XOR (or CNOT) gate is
realized, cluster states (and consequently, a one-way quantum computer)
can be generated. This result does not depend on the type of
interaction in the system. Therefore, other proposals such as
trapped ion \cite{Cirac} and superconducting qubits \cite{Moij}, can be
used as well, to create  cluster states.

\section{Concluding Remarks}

In summary,
an alternative way, using Heisenberg interaction between qubits, was
introduced to create cluster states which is useful for solid
state systems. In this method the qubits in the cluster are entangled pairwise,
leading to $2d$ steps in $d$-dimensional cubic lattices. Furthermore, by tuning
the coupling strengths of the interaction, it is possible to create
cluster states via anisotropic Heisenberg exchange interaction. Experimentally, these
cluster states can be generated in coupled quantum dots or similar systems.

\section{Acknowledgment}

We are grateful to W. A. Coish, H. Gassmann
and J. Schliemann for helpful discussions. This work has been supported
by the Swiss NSF,  NCCR Nanoscience, DARPA, and ARO.\\


\begin{references}

\bibitem{niel}
M. A. Nielsen and I. L. Chuang {\textit{Quantum Computation and
Quantum Information}} (Cambridge University Press, New York, 2000).


\bibitem{br}
H.-J. Briegel and R. Raussendorf, Phys. Rev. Lett. \(\textbf{{86}}\), 910
(2001).


\bibitem{hu}
K. Huang, {\textit{Statistical Mechanics}} , second edition (John
Wiley, Singapore, 1987).


\bibitem{rauss}
R. Raussendorf and H.-J. Briegel, Phys. Rev. Lett. \(\textbf{{86}}\), 5188
(2001).


\bibitem{rbb} 
R. Raussendorf, D. E. Browne, and H.-J. Briegel, 
Phys. Rev. A \(\textbf{{68}}\), 022312 (2003).


\bibitem{ash}
N. W. Ashcroft and N. D. Mermin, {\textit{Solid State Physics}}
(Saunders, Philadelphia, 1976), Chap. 32.

\bibitem{loss}
D. Loss and D. P. DiVincenzo, Phys. Rev. A \(\textbf{{57}}\), 120 (1998).


\bibitem{ising}
For the special case $n=m=0$, the time evolution operator
$U_{SAH}^{(a, a+1)}$  reduces to Eq. (\ref{U1}), up to a minus sign,
depending on the value of $k$.


\bibitem{cycle}
$x,\, y$ and $z$ can be changed in cyclic permutation.


\bibitem{bur}
G. Burkard, D. Loss, and D. P. DiVincenzo, Phys. Rev. B \(\textbf{{59}}\),
2070 (1999).


\bibitem{awschalom}
For a review see G. Burkard and D. Loss,  pp. $229$, in
{\it Semiconductor Spintronics and Quantum Computation},
D.~D. Awschalom, D. Loss, and N. Samarth (eds.), (Springer, Berlin, 2002).


\bibitem{bar}
A. Barenco \(et\,\,al\). Phys. Rev. A \(\textbf{{52}}\), 3457 (1995).



\bibitem{Smolin}
G. Burkard, D. Loss, D.P. DiVincenzo, and J.A. Smolin, Phys. Rev. B
\(\textbf{{60}}\), 11404 (1999).


\bibitem{Schl}
J. Schliemann, D. Loss, and A.H.  MacDonald, 
Phys. Rev. B, \(\textbf{{63}}\), 085311 (2001).


\bibitem{Req}
R. Requist, J. Schliemann, A.G. Abanov, and  D. Loss,
cond-mat/0409096.  

\bibitem{Cirac}
J.I. Cirac and P. Zoller,
Phys. Rev. Lett. \(\textbf{{74}}\), 4091
(1995).

\bibitem{Moij}
J.E. Moij  \(et\,\,al\). Science,
\(\textbf{{285}}\), 1036 (1999).

\end{references}
\end{document}